\documentclass[reprint,twocolumn,superscriptaddress,showpacs,preprintnumbers,nofootinbib,aps,prd,
]{revtex4-1}

\usepackage{caption}
\usepackage{graphicx,color}
\usepackage{amsmath,amssymb,amsfonts}
\usepackage{stackrel}
\usepackage{enumitem}
\usepackage[english]{babel}
\usepackage{verbatim}
\usepackage{comment}
\usepackage{xcolor}
\usepackage{subfigure}
\usepackage{dcolumn}
\usepackage{physics}
\usepackage{units}
\usepackage{natbib}
\usepackage[section]{placeins}
\usepackage[normalem]{ulem} 

\usepackage{eso-pic}

\usepackage[colorlinks=true,
            linkcolor=red,
            urlcolor=blue,
            citecolor=blue]{hyperref}
\graphicspath{
  {./}
  {figures/}
}

\newcommand{\tf}{\texorpdfstring}
\newcommand{\gev}{~\text{GeV}}
\newcommand{\tev}{~\text{TeV}}

\def\nn{\nonumber}

\newcommand{\abi}{~\text{ab}^{-1}}
\newcommand{\fbi}{~\text{fb}^{-1}}

\newcommand{\dcs}{doubly-charged scalar }
\newcommand{\dR}{\Delta_R^{\pm\pm}}

\definecolor{orange}{rgb}{1,0.5,0}
\definecolor{amethyst}{rgb}{0.6, 0.4, 0.8}
\definecolor{antiquefuchsia}{rgb}{0.57, 0.36, 0.51}
\definecolor{byzantine}{rgb}{0.74, 0.2, 0.64}
\definecolor{blue-violet}{rgb}{0.54, 0.17, 0.89}
\definecolor{cadmiumred}{rgb}{0.89, 0.0, 0.13}
\definecolor{brightcerulean}{rgb}{0.11, 0.67, 0.84}

\begin{document}

\setlength{\abovedisplayskip}{6pt}
\setlength{\belowdisplayskip}{6pt}

\preprint{CTPU-PTC-25-43}

\title{Probing flavor-diagonal couplings of doubly-charged scalar at low and high energies}
\author{Gang Li}
\email{ligang65@mail.sysu.edu.cn (corresponding author)}
\affiliation{School of Physics and Astronomy, Sun Yat-sen University, Zhuhai 519082, China}
\affiliation{
Guangdong Provincial Key Laboratory of Quantum Metrology and Sensing, Sun Yat-Sen University, Zhuhai 519082, China}

\author{Jin Sun}
\email{sunjin0810@ibs.re.kr (corresponding author)}
\affiliation{
Particle Theory and Cosmology Group, Center for Theoretical Physics of the Universe,
Institute for Basic Science (IBS), Daejeon 34126, Korea}

\begin{abstract}

We investigate the phenomenology of a TeV-scale doubly-charged scalar from the right-handed sector within the framework of left-right symmetric models. Focusing on its flavor-diagonal couplings to right-handed electrons and muons, we assess probes from both high-energy colliders and low-energy precision experiments. High-energy processes include Bhabha scattering at LEP and future circular electron-positron colliders (CEPC/FCC-ee), direct production at the LHC, and dedicated searches and precision measurements at proposed muon colliders and $\mu$TRISTAN. Low-energy observables encompass parity-violating Møller scattering, muon anomalous magnetic moment, and muonium-antimuonium oscillations. Our combined analysis indicates that for a doubly-charged scalar in the $1$-$3$ TeV range, the flavor-diagonal Yukawa couplings to electrons and muons as small as $10^{-2}$ are accessible. Observations of such a doubly-charged scalar would potentially point toward the type-I seesaw mechanism of neutrino masses in the left-right symmetric model with $D$-parity breaking.

\end{abstract}

\maketitle

\section{Introduction}

The origin of neutrino masses remains one of the most profound open questions in particle physics. Explaining the tiny neutrino masses requires particles beyond the Standard Model (BSM). For example, in the type-I seesaw models~\cite{Minkowski:1977sc, Yanagida:1979as, Gell-Mann:1979vob, Mohapatra:1979ia, Glashow:1979nm}, right-handed (RH) neutrinos are introduced, while 
in the type-II seesaw models~\cite{Konetschny:1977bn,Magg:1980ut,Schechter:1980gr,Mohapatra:1980yp,Cheng:1980qt}, left-right symmetric models (LRSMs)~\cite{Pati:1974yy,Mohapatra:1974gc,Mohapatra:1974hk,Senjanovic:1975rk,Senjanovic:1978ev}, radiative neutrino mass models~\cite{Cheng:1980qt,Zee:1985id,Babu:1988ki} and $d=7$ neutrino mass models~\cite{Babu:2009aq,Bonnet:2009ej} include
one or more doubly-charged scalars. 
Seeking for new particles via their interactions with charged leptons thus paves one of the pathways towards 
verifying these scenarios.

In general, the Yukawa couplings of \dcs are connected to neutrino masses and mixing, and contribute to charged lepton flavor violating (CLFV) processes.
If the \dcs is an $SU(2)$ singlet~\cite{Cheng:1980qt,Zee:1985id,Babu:1988ki}, it directly induces neutrino masses and lepton flavor mixing at the loop levels. If it arises from a scalar field in a higher $SU(2)$ representation, however, the Yukawa couplings of the \dcs are the same as those of neutral component of the scalar field, which is responsible for generating neutrino masses and lepton flavor mixing.
For example, in the minimal LRSM~\cite{Mohapatra:1979ia,Mohapatra:1980yp}, doubly-charged scalars emerge from the triplet scalars $\Delta_{L,R}$ after the spontaneous symmetry breaking of $SU(2)_L \times SU(2)_R \times U(1)_{B-L}$ gauge group. Since the left-right symmetry is explicitly imposed in the Yukawa sector, their couplings to leptons are equal and thus both severely constrained by the CLFV searches~\cite{Das:2012ii,Barry:2013xxa,Davidson:2022jai}.

However, in non-manifest LRSMs, the RH Yukawa couplings of the \dcs may differ from those of the left-handed ones. In particular, in LRSMs with $D$-parity breaking~\cite{Chang:1983fu,Chang:1984uy}, parity and the $SU(2)_R$ breaking scale decouple, allowing the mass of RH \dcs $\Delta_R^{\pm\pm}$ to be considerably 
smaller than the RH scale $v_R$. Within this LRSM framework, neutrino masses can be generated via either the type-I seesaw mechanism~\cite{Chang:1984uy} or the type-II seesaw mechanism~\cite{Sahu:2006pf,Deppisch:2014zta}.

Refs.~\cite{Dev:2018sel,Li:2024djp} investigated the sensitivities to the coupling of $\Delta_R^{\pm\pm}$ to electrons in both low- and high-energy processes, assuming a $\Delta_R^{\pm\pm}$ mass at the TeV scale, consistent with existing constraints from direct searches at the Large Hadron Collider (LHC). Recently, Ref.~\cite{Akhmedov:2024rvp} proposed a concrete LRSM featuring a long-lived $\Delta_R^{\pm\pm}$. In this scenario, if the Yukawa couplings to charged leptons satisfy $|f_R| \lesssim 10^{-8}$, TeV-scale $\Delta_R^{\pm\pm}$ can decay outside the detector volume. Nevertheless, the magnitudes of $(f_R)_{\alpha\beta}$ are treated as being of the same order across different flavor indices $\alpha, \beta$, so that the RH neutrino masses $M_R \sim f_R v_R$ are too small to generate light neutrino masses via the type-I seesaw mechanism, necessitating the introduction of additional fields~\cite{Akhmedov:2024rvp}.

In this work, we investigate the sensitivities to the RH Yukawa couplings $f_R^{\alpha\beta}$ for $\alpha, \beta = e, \mu$ in various low- and high-energy processes. 
Different from the parameter regions considered in Ref.~\cite{Akhmedov:2024rvp}, we focus on scenarios where the RH \dcs $\Delta_R^{\pm\pm}$ resides at the TeV scale and decays promptly. We further assume that the flavor off-diagonal couplings are significantly suppressed relative to the flavor-diagonal ones, which could be realized by invoking flavor symmetries~\cite{Rodejohann:2015hka}, thereby satisfying stringent constraints from CLFV searches and allowing for sizable lepton-flavor-conserving coupling $f_R^{ee}$ or $f_R^{\mu\mu}$. This assumption is justified since, light neutrino masses in the type-I seesaw mechanism are given by $M_\nu = -M_D M_R^{-1} M_D^T$, where the Dirac neutrino mass matrix $M_D$ is generally complex. If future experiments observe no CLFV signals but detect signals in charged lepton-flavor-conserving processes, such observations could potentially point toward the RH \dcs in non-manifest LRSMs, possibly the LRSM with $D$-parity breaking. For recent studies on a TeV-scale \dcs focusing on lepton-flavor-violating observables, see Refs.~\cite{Crivellin:2018ahj,Dev:2021axj,Dev:2023nha}.

The remainder of this paper is organized as follows. 
In Sec.~\ref{sec:model}, we introduce the minimal model for the doubly-charged scalar that interacts with the leptons.
We investigate the low-energy observables and high-energy processes that can probe the mass and Yukawa couplings of the doubly-charged scalar in Sec.~\ref{sec:low} and Sec.~\ref{sec4}, respectively. In Sec.~\ref{sec:discussion}, we present the combined results and discuss their implications. We conclude in Sec.~\ref{sec:conclusion}.

\section{Minimal model}
\label{sec:model}

We consider a minimal model motivated by the LRSM with $D$-parity breaking~\cite{Chang:1983fu,Chang:1984uy} (see Ref.~\cite{Akhmedov:2024rvp} for recent studies), which is based on the gauge group $SU(2)_L \times SU(2)_R \times U(1)_{B-L}$,
with $B$ and $L$ being the baryon and lepton numbers, respectively. The relevant Yukawa interactions of triplet scalar to leptons are given by
\begin{align}
\label{eq:yuk}
\mathcal{L}_Y  
&= - \left( \bar{L}_L^c i\tau_2 \Delta_L f_L L_L +  \bar{L}_R^c i\tau_2 \Delta_R f_R L_R\right) + {\rm h.c.}\;,
\end{align}
where
the LH and RH lepton doublets are defined as
\begin{align}
L_{L}=\left(\begin{array}{l}
\nu_L \\
\ell_L
\end{array}\right)\;, \quad L_{R}=\left(\begin{array}{c}
\nu_R \\
\ell_R
\end{array}\right)\;,
\end{align}
and the scalar triplets are
\begin{align}
\Delta_{L, R}=\left(\begin{array}{cc}
\Delta_{L, R}^{+} / \sqrt{2} & \Delta_{L, R}^{++} \\
\Delta_{L, R}^0 & -\Delta_{L, R}^{+} / \sqrt{2}
\end{array}\right)\;.
\end{align}
In Eq.~\eqref{eq:yuk}, $\tau_2$ is the second Pauli matrix, $f_{L,R}$  represents the Yukawa coupling matrices, 
and $\bar L_{L,R}^c \equiv L_{L,R}^T C$ with $C$ being the charge conjugation matrix, 
and ``h.c.'' denotes the Hermitian conjugate.

After spontaneous symmetry breaking, the neutral components of $\Delta_{L,R}$ acquire vacuum expectation values (vevs), $\langle \Delta_{L,R}^0 \rangle = v_{L,R} / \sqrt{2}$. The masses of $\Delta_L$ are 
expected to be at the $D$-parity breaking scale, which lies significantly above the $SU(2)_R$ breaking scale~\cite{Chang:1983fu,Chang:1984uy}, thus $\Delta_L$ effectively decouples from the processes considered here. 
It was emphasized in Ref.~\cite{Akhmedov:2024rvp} that the mass of the singly-charged scalar $H^{\pm}$ must exceed 15~TeV to satisfy constraints from $K-\bar{K}$ and $B-\bar{B}$ mixings~\cite{Zhang:2007da,Bertolini:2014sua,Bertolini:2019out,Dekens:2021bro}, whereas the mass of the \dcs $\Delta_R^{\pm\pm}$ can be arbitrarily small.
This is because
$H^\pm$ are the physical singly-charged mass eigenstates built from the charged components of the bidoublet scalar $\Phi$, while $\Delta_R^\pm$ are the would-be Goldstone bosons eaten by $W_R^\pm$.
The squared masses of $H^\pm$ and $\Delta_R^{\pm\pm}$ are given by
\begin{align}
m^2_{H^\pm}\simeq \alpha_3 (1+2\xi^2) v_R^2/2\;,\quad m^2_{\Delta_R^{\pm\pm}}\simeq 2\rho_2 v_R^2\;,
\end{align}
where $\alpha_3$ and $\rho_2$ are parameters of the scalar potential: $\alpha_3$ governs the mixing between the bidoublet $\Phi$ and triplet $\Delta_R$, $\rho_2$ represents the quartic coupling of $\Delta_R$, and $\xi$ denotes the ratio of the two vevs of $\Phi$. Further details can be found in Refs.~\cite{Akhmedov:2024rvp,Dev:2016dja,Kriewald:2024cgr}.
In addition, we assume that the vev $v_R$ is sufficiently large, so that the RH gauge bosons are also irrelevant to our analysis, which focuses on $\Delta_R^{\pm\pm}$ in both low- and high-energy processes\,\footnote{Consequently, we do not consider the searches for neutrinoless double beta decay. For a detailed study of the scenarios where $W_R$ does not decouple, see Ref.~\cite{Li:2024djp}.}.

Furthermore, we assume the flavor off-diagonal Yukawa couplings $f_R^{\alpha\beta}$ with $\alpha \neq \beta$ are negligible, while the diagonal couplings satisfy $|f_R^{\alpha\alpha}| \geq 10^{-3}$ for $\alpha = e,\mu$. Under this assumption, viable light neutrino masses and mixing can still be achieved in the type-I seesaw mechanism without the need of introducing additional singlet fermion fields, as required in Ref.~\cite{Akhmedov:2024rvp}.
This is possible because the RH neutrino masses $M_R \sim f_R v_R$ could have adequate magnitude, and the Dirac neutrino mass matrix $M_D$ is a general complex matrix.  
For simplicity, we take the leptonic Yukawa couplings
to be real and positive\,\footnote{Relaxing this assumption would introduce additional CP-violating phases that contribute to lepton electric dipole moments~\cite{Liao:2008ix}, and generate CP asymmetry required for leptogenesis through heavy neutrino decays~\cite{ODonnell:1993obr}. } throughout this work.

\section{Low-energy observables}
\label{sec:low}

In this section, we will investigate constraints on the mass and Yukawa couplings of $\dR$ using the low-energy observables, including parity-violating asymmetry in Møller scattering, muon anomalous magnetic momentum, and muonium-antimuonium transition probability.

\subsection{Møller scattering}

It has been investigated in Refs.~\cite{Rizzo:1981dla,Dev:2018sel} that the couplings of $\Delta_R^{\pm\pm}$  to the electrons can be probed in the Møller scattering ($e^- e^- \to e^- e^-$) process via the $s$-channel exchange of $\Delta_R^{\pm\pm}$. By using the Fierz transformation,
we derive the effective interaction at low energies as
\begin{align}
{\cal L}_{\rm PV} = \frac{(f_{R}^{ee})^2}{2m_{\Delta^{++}}^2}(\bar{e}_R\gamma^\mu e_R)(\bar{e}_R\gamma_\mu e_R)  \,,
\label{eq:L_PV}
\end{align}
where we have ignored the contribution from the left-handed doubly-charged scalar, and $m_{\Delta^{++}}$ represents the mass of the RH \dcs $\Delta_R^{\pm\pm}$ for brevity. 
The upcoming MOLLER experiment~\cite{MOLLER:2014iki}, which aims to measure the parity-violating asymmetry of the Møller scattering with unprecedented sensitivity, can put the following low bound
\begin{align}
    \dfrac{m_{\Delta^{++}}}{f_R^{ee}} > 7.6\tev\;
\end{align}
at $95\%$ confidence level (C.L.)~\cite{Li:2024djp}. 

\subsection{Muon \tf{$g-2$}{g-2}}

Given the coupling to muons $f_R^{\mu\mu}$, the \dcs $\Delta_R^{\pm\pm}$ can contribute to the muon anomalous magnetic moment, i.e., $(g-2)_\mu$, 
at one-loop level, which is expressed as~\cite{Leveille:1977rc,Gunion:1989in,Abada:2007ux,Cheng:2021okr,Huang:2024iip} 
\begin{eqnarray}\label{largeg-2}
\Delta a_{\mu}
= -\frac{m_{\mu}^2}{24\pi^2}
   \frac{(f_R^{\mu\mu})^2}{m_{\Delta^{++}}^2}\;,
\end{eqnarray}   
with $m_\mu$ being the muon mass.
Based on the latest measurement of $(g-2)_\mu$ by the FNAL experiment~\cite{Muong-2:2025xyk},
and the updated SM prediction in lattice-QCD calculation~\cite{Aliberti:2025beg},  
we obtain the difference 
\begin{align}
   a_\mu^{\rm exp}-a_\mu^{\rm SM}=39(64)\times 10^{-11}\;,
\end{align}
which
can be negative at $1\sigma$ level.
We thus obtain an lower bound on 
$m_{\Delta^{++}}/f_R^{\mu\mu} > 0.23\tev$
at $2\sigma$ level from the $(g-2)_\mu$ measurement.

Similarly, the coupling of $\Delta_R^{\pm\pm}$ to electrons yields a negative contribution to the electron anomalous magnetic moment $(g-2)_e$. However,  depending on whether the fine-structure constant 
is determined from rubidium~\cite{Morel:2020dww} or cesium~\cite{Parker:2018vye}, 
the difference between the experimental value and the SM prediction currently shows a sign discrepancy~\cite{Gabrielse:2025jep}. It remains premature to draw definitive conclusions. Therefore, we do not include constraints from $(g-2)_e$ in this work.

\subsection{Muonium-antimuonium transition}

The muonium to antimuonium transition has recently drawn considerable attention both theoretically~\cite{Conlin:2020veq,Han:2021nod,Fukuyama:2021iyw,Fukuyama:2022dhe,Fukuyama:2023drl,Huang:2025dga,Ghosh:2025oju} and experimentally~\cite{Bai:2022sxq,Bai:2024skk,Kawamura:2021lqk}.
The muonium (dubbed $M$) is a bound state of $\mu^+$ and $e^-$, thus
this process can probe BSM physics that violates lepton flavor number by two units $\Delta L_\mu = - \Delta L_e = 2$~\cite{Feinberg:1961zza,Conlin:2020veq,Heeck:2024uiz}, which is not directly constrained by the CLFV searches with $\Delta L = 1$~\cite{Davidson:2022jai}.

Besides, $M-\overline{M}$ transition can also be induced by $t$-channel exchange of a doubly-charged scalar~\cite{Chang:1989uk,Han:2021nod,Fukuyama:2021iyw}, without involving lepton flavor violation. At low energies, we obtain the following effective interaction~\cite{Swartz:1989qz,Chang:1989uk}
\begin{eqnarray}
  \mathcal{L}_{M-\overline M}=\frac{f_R^{ee} f_R^{\mu\mu}}{2m^2_{\Delta^{++}}}  (\bar \mu_R \gamma^\mu e_R)(\bar \mu_R \gamma^\mu e_R)\;.
\end{eqnarray}

Depending on the spin orientation, muonium can be produced in either the spin-0 or spin-1 state, referred to as para-muonium $(M_P)$ and ortho-muonium $(M_V)$, respectively. The total transition probaliblity is a weighted sum of these two contributions~\cite{Conlin:2020veq}
\begin{align}
    P(M \to \overline{M}) = \sum_{i=P,V} f_i P(M_i \to \overline{M}_i) \;,
\end{align}
where the fractions $f_P + f_V = 1$.
Following Refs.~\cite{Chang:1989uk,Cvetic:2005gx}, we parameterize the transition probability of para-muonium as
\begin{align}
    P(M_P \to \overline{M}_P) &=64^3(\frac{3\pi^2\alpha_{\rm em}^3}{G_Fm_\mu^2})^2(\frac{m_e}{m_\mu})^6(\frac{G_{M\overline{M}}}{G_F})^2\;\nn\\
    &= 2.64\times 10^{-5} (\frac{G_{M\overline{M}}}{G_F})^2\;,
\end{align}
where $\alpha_{\rm em}$ is the fine-structure constant, $m_e$ is the electron mass, and $G_F$ denotes the Fermi constant. The effective coupling is given by~\cite{Chang:1989uk,Conlin:2020veq,Han:2021nod}
\begin{align}
    G_{M\overline{M}} = \dfrac{f_R^{ee} f_R^{\mu\mu}}{4\sqrt{2} m^2_{\Delta^{++}}}\;.
\end{align}
The transition probability of otho-muonium is~\cite{Conlin:2020veq}
\begin{align}
    P(M_V \to \overline{M}_V) = 9 P(M_P \to \overline{M}_P)\;.
\end{align}

The most stringent constraint to date comes from the MACS experiment at PSI~\cite{Willmann:1998gd}, which places an upper limit on the $M -\overline{M}$ transition probability of $P(M \to \overline{M}) < 8.3 \times 10^{-11}/S_B(B_0)$ at 90\% confidence level (CL).
The factor $S_B(B_0)$ accounts for the suppression of the transition due to the external magnetic field~\cite{Horikawa:1995ae,Hou:1995np}. For the MACS experiment, the magnetic field strength $B_0 = 0.1\,\mathrm{T}$, leading to $S_B(B_0) = 0.35$~\cite{Willmann:1998gd}. The proposed MACE experiment, 
operating at the same magnetic field strength, is expected to improve the sensitivity beyond the level of $10^{-13}$~\cite{Bai:2024skk}.
From Ref.~\cite{Fukuyama:2021iyw}, we can effectively take $f_p = 1$ for the MACS experiment, and derive the upper bound $G_{M\overline{M}}/G_F < 3 \times 10^{-3}$. On the other hand, as in Ref.~\cite{Bai:2024skk}, we estimate the sensitivity of MACE by assuming $f_P = 0.3$, and obtain 
$G_{M\overline{M}}/G_F < 4 \times 10^{-5}$. 

\bigskip

\section{High-energy processes}
\label{sec4}

The \dcs can also contribute to high-energy processes at hadron and lepton colliders. 
The most stringent constraint on the mass of the doubly-charged scalar $\Delta_R^{\pm\pm}$ arises from direct searches conducted at the LHC, specifically through their pair production in the Drell-Yan process $pp\to \Delta_R^{++}\Delta_R^{--}$. Each scalar decays into a pair of same-sign leptons. Based on the searches performed during LHC Run 2, which analyzed an integrated luminosity of $139\fbi$, the mass range $m_{\Delta_R^{++}} < 1.08 \tev$ has been excluded~\cite{ATLAS:2022pbd}. We adopt a conservative lower limit of $m_{\Delta_R^{++}} \geq 1.4\ \text{TeV}$, since the decay branching ratios of $\Delta_R^{\pm\pm}$  are different from those assumed in the experimental analysis.

On the other hand, bound on ${f_R^{ee}}/{m_{\Delta^{++}}}$ is set by Bhabha scattering $e^+e^-\to e^+e^-$ at the Large Electron-Positron Collider (LEP). Reinterpreting the combined LEP limit
from data collected at DELPHI, ALEPH, and OPAL~\cite{ALEPH:2013dgf}, which correspond to the mean center-of-mass energy $\sqrt{s}=195.6\gev$ and a total integrated luminosity of $\mathcal{L} = 745~\text{pb}^{-1}$~\cite{DELPHI:2005wxt}, we obtain~\cite{Nomura:2017abh,Li:2024djp}
\begin{align}
\dfrac{m_{\Delta^{++}}}{f_R^{ee}} > 2.43\tev\;.
\end{align}

\begin{figure*}
    \centering
    \subfigure[\label{fig:eeee}]
 	{\includegraphics[width=.3\textwidth]{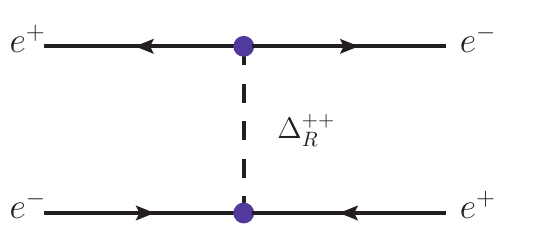}}
    \subfigure[\label{fig:mupmummumu}]
 	{\includegraphics[width=.3\textwidth]{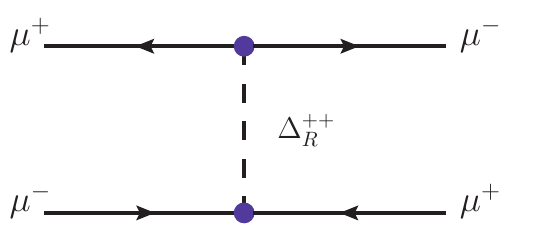}}
    \subfigure[\label{fig:deltadelta}]
 	{\includegraphics[width=.3\textwidth]{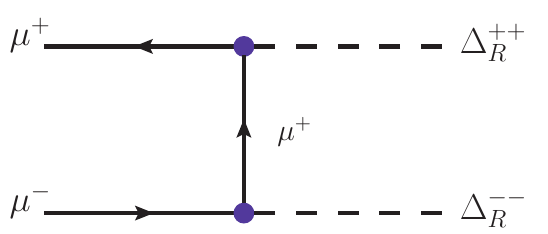}}
     \subfigure[\label{fig:mumumumu}]
 	{\includegraphics[width=.3\textwidth]{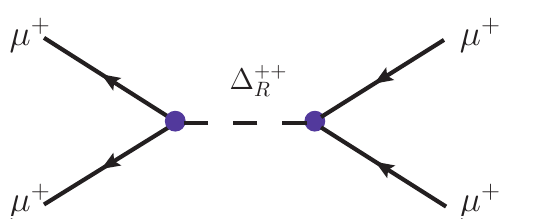}}
     \subfigure[\label{fig:mumuee}]
     {\includegraphics[width=.3\textwidth]{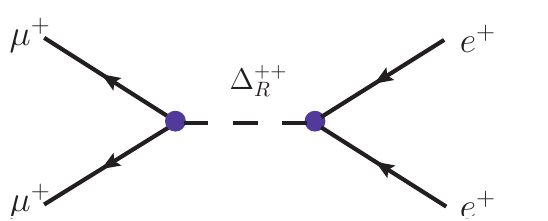}}
     \subfigure[\label{fig:deltagamma1}]
 	{\includegraphics[width=.3\textwidth]{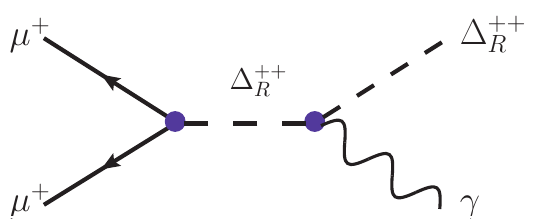}}
     \subfigure[\label{fig:deltagamma2}]
 	{\includegraphics[width=.3\textwidth]{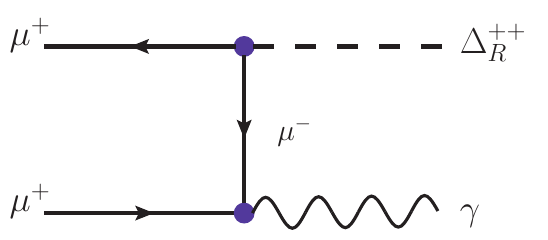}}
    \subfigure[\label{fig:muemue}]
 	{\includegraphics[width=.3\textwidth]{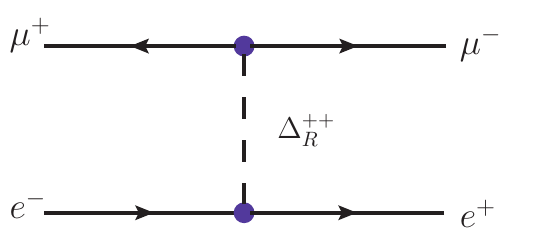}}
    \caption{
    Representative Feynman diagrams for
    the high-energy processes with the exchange or production of $\Delta^{\pm\pm}_R$.
    }
    \label{fig:feynman}
\end{figure*}

It is evident that the current constraints on Yukawa couplings to leptons are relatively weak, particularly in light of the mass limits on the \dcs established by direct searches at the LHC. In the below, we will examine the sensitivities of the processes illustrated in Fig.~\ref{fig:feynman} to the ratio $f_R^{\alpha\alpha}/m_{\Delta^{++}}$ at future high-energy lepton colliders\,\footnote{Similar processes have been proposed long time ago~\cite{Rizzo:1981dla} to search for doubly-charged scalars in the LRSM at $e^+e^-/e^-e^-$ colliders.}.

\begin{itemize}
\item $e^+e^-$ colliders:
CEPC~\cite{CEPCStudyGroup:2018ghi} and FCC-ee~\cite{FCC:2018evy} with $\sqrt{s}=240\gev$ and $\mathcal{L} = 5\abi$.
\item $\mu^+\mu^-$ colliders: future muon colliders~\cite{MuonCollider:2022nsa,InternationalMuonCollider:2024jyv} 
with $\sqrt{s}=3\tev\,(10\tev)$ and $\mathcal{L} = 1\abi\,(10\abi)$.
\item $\mu^+\mu^+$ collider:  
$\mu$TRISTAN with $\sqrt{s}=2\tev$ and $\mathcal{L} = 12\fbi$
~\cite{Hamada:2022mua,Hamada:2022uyn}.
\item $\mu^+e^-$ collider: 
$\mu$TRISTAN with $\sqrt{s}=346\gev$ and $\mathcal{L} = 1\abi$~\cite{Hamada:2022mua,Hamada:2022uyn}.
\end{itemize}

\begin{figure}[!htb]
    \centering
\includegraphics[width=0.45\textwidth]{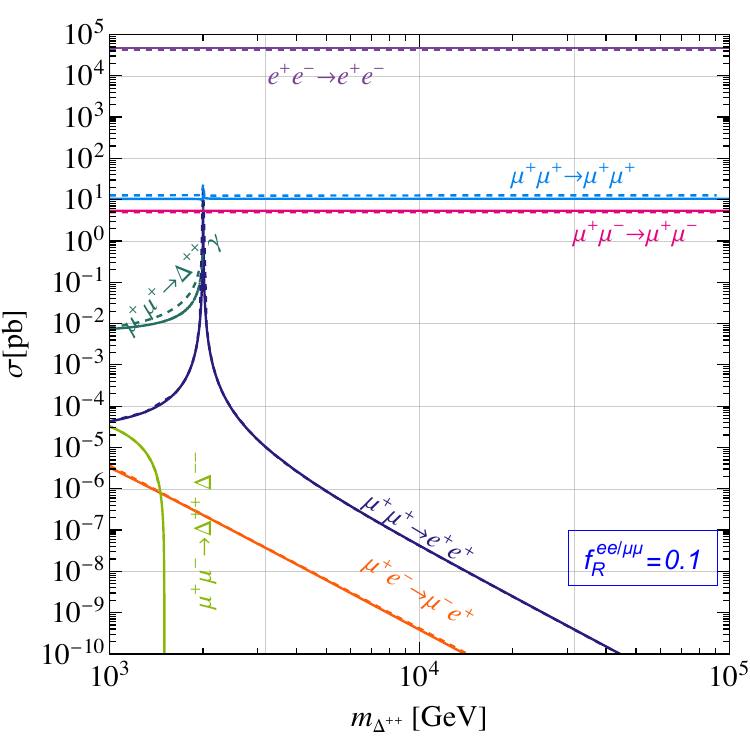}
    \caption{
    Total cross section as a function of $m_{\Delta^{++}}$ for various processes at different colliders: 
    $e^+e^-$ ($\sqrt{s} = 240$~GeV), $\mu^+\mu^-$ ($\sqrt{s} = 3\tev$), $\mu^+\mu^+$ ($\sqrt{s} = 2$~TeV), and $\mu^+e^-$ ($\sqrt{s} = 346$~GeV). Analytical results are shown as solid lines; cross-checks from \textsc{MadGraph5\_aMC@NLO} are shown as dashed lines. 
    }
    \label{fig:cross}
\end{figure}

The solid curves in Fig.~\ref{fig:cross} show the analytically computed cross sections versus $m_{\Delta^{++}}$ for the benchmark values $f_R^{ee} = f_R^{\mu\mu} = 0.1$. They are cross-checked with \textsc{MadGraph5\_aMC@NLO}~\cite{Alwall:2014hca,Frederix:2018nkq} (dashed curves). The results agree well with each other.

For convenience, we separate the total cross section 
\begin{align}
\label{eq:xsec}
    \sigma_i = \sigma_i^{\rm SM} + \sigma_i^{\rm NP}\;,
\end{align}
where $i$ represents the process under consideration,  the terms with superscripts ``SM'' and ``NP'' denotes the contributions from the SM and doubly-charged scalar including its interference with the SM, respectively.

It should be emphasized that in this work we do not consider polarized initial beams and differential distributions at these future lepton colliders. Nevertheless, it has been shown that polarized muon beams can be used to distinguish $\Delta_R^{\pm\pm}$ from $\Delta_L^{\pm\pm}$ via pair production, if both states are kinematically accessible, since the production rates exhibit a pronounced dependence on the beam polarization due to their different chiral couplings to leptons~\cite{Belfkir:2023lot}. Furthermore, with polarized lepton beams, the angular distributions of lepton-pair production processes become sensitive to the chiral structure of the Yukawa interactions in Eq.~\eqref{eq:yuk}, resulting in different forward-backward asymmetries~\cite{Nomura:2017abh}.

\subsection{CEPC/FCC-ee}

At future electron-positron colliders CEPC/FCC-ee~\cite{CEPCStudyGroup:2018ghi,FCC:2018evy} with the center-of-mass energy $\sqrt{s}=240\gev$, Bhabha scattering is expected to be measured with higher precision than the LEP due to a larger number of signal events.
In order to derive the sensitivities of CEPC/FCC-ee to ${m_{\Delta^{++}}}/{f_R^{ee}}$, we calculate the cross section of Bhabha scattering involving the contribution from $\Delta_R^{\pm\pm}$.

The Bhabha scattering $e^+e^-\to e^+e^-$  occurs via the $s$-channel and $t$-channel exchange of $\gamma/Z$ in the SM. The \dcs $\Delta_R^{\pm\pm}$ can contribute to this process in the $u$ channel, as depicted in Fig.~\ref{fig:eeee}. 

We obtain
\begin{align}
    \sigma_{e^+e^-\to e^+ e^-}^{\rm NP} =\frac{1}{16\pi s^2} \int_{t_{\rm min}}^{t_{\rm max}} {\rm d} t \left[ \mathcal |\mathcal M_{\Delta}|^2 + \mathcal{M}^2_{\rm int} \right]\;,
\end{align}
where the squared amplitude for the exchange of $\dR$ and the term describing interference with the SM are expressed as
\begin{eqnarray}
  |\mathcal M_{\Delta}|^2&=\dfrac{(f_R^{ee})^4 u^2}{4 [(u-m_{\Delta^{++}}^2)^2+\Gamma_\Delta^2 m_{\Delta^{++}}^2]}\;,\\
   \mathcal{M}^2_{\rm int}&=\dfrac{ e^2 (f_R^{ee})^2 u^2[g(s)-g(t)]}{4 [(u-m_{\Delta^{++}}^2)^2+\Gamma_\Delta^2 m_{\Delta^{++}}^2]}\;.
 \end{eqnarray} 
Hereafter, the lepton masses are neglected.
The function $g(x)$ is defined as
 \begin{align}
  g(x)&=(u-m_{\Delta^{++}}^2)\left[\frac{ 4}{x}+
   \frac{ \left(x- m_Z^2\right)
   }{4 c_W^2 s_W^2 \left|D_Z(x)\right|^2}\right]\nn\\
   &\quad+\frac{m_{\Delta^{++}} \Gamma_\Delta m_Z \Gamma_Z}{4 c_W^2 s_W^2
   \left|D_Z(x)\right|^2}\;, 
\end{align}
with 
\begin{align}
    D_Z(x) \equiv x-m_Z^2 + i \Gamma_Z m_Z\;.
\end{align}
In the above, $\Gamma_Z$ and $m_Z$ are the width and mass of $Z$ boson, respectively, while $\Gamma_\Delta$ denotes the width of $\dR$. 
Besides, $s$, $t$ and $u$ are the Mandelstam variables. $s_W$ and $c_W$ denote the sine and cosine of the weak mixing angle, respectively. 

The variable $t$ is a monotonic function of the pseudo-rapidity $\eta$,
\begin{align}
\label{eq:rapidity}
    t=-\frac{s}{1+e^{2\eta}}\;,\quad \eta \equiv -\ln\tan(\theta/2)\;.
\end{align}
In experimental searches, cut on the rapidity $|\eta| < \eta_{\rm cut}$ should be imposed,  which leads to
\begin{align}
\label{eq:tmin}
    t_{\rm min}=-\frac{s}{1+e^{-2\eta_m}}\, \quad
t_{\rm max}=-\frac{s}{1+e^{2\eta_m}}\;.
\end{align}

The measured cross section at LEP~\cite{ALEPH:2013dgf} required the average polar scattering angle to satisfy $|\cos\theta| < 0.90$ ($|\eta| < 1.47$). In contrast, detectors at future circular colliders like CEPC/FCC-ee can cover a larger angular range. 
In practice, we take $|\cos\theta| < 0.9998$ ($|\eta| < 4.56$) based on Ref.~\cite{Sun2025Lumical} for the studies CEPC/FCC-ee~\cite{CEPCStudyGroup:2018ghi,FCC:2018evy}.

The purple curve in Fig.~\ref{fig:cross} shows the total cross section $\sigma_{e^+ e^- \to e^+ e^-}$ for Bhabha scattering at CEPC/FCC-ee for $f_{R}^{ee} = 0.1$. This cross section is larger than that at LEP. The increase occurs because the differential cross section peaks in the forward direction ($|\cos\theta| \to 1$). The CEPC/FCC-ee's enhanced angular acceptance captures more of these forward events. This compensates for the kinematic suppression from the higher center-of-mass energy and results in a larger total cross section.

Besides, we can see that the cross section only has a mild dependence on $m_{\Delta^{++}}$. For example, for $m_{\Delta^{++}}=1.4\tev$, we obtain $\sigma_{e^+e^-\to e^+ e^-}^{\rm NP}/\sigma_{e^+e^-\to e^+ e^-} = - 7\times 10^{-7}$. It indicates that interference with the SM is destructive and $\sigma_{e^+e^-\to e^+ e^-}^{\rm NP}$ approximately scales with $\left(f_R^{ee}/m_{\Delta^{++}}\right)^2$.

To evaluate the sensitivity, we derive the exclusion at 95\% confidence level (CL)  using the following criterion
\begin{align}
\label{eq:crit}
    \dfrac{n_s}{\sqrt{n_s + n_b}} \geq 1.96\;.
\end{align}

For the Bhabha scattering at CEPC/FCC-ee, the numbers of events are given by
\begin{align}
\label{eq:events_CEPC}
    n_s = -\sigma_{e^+e^-\to e^+ e^-}^{\rm NP} \mathcal{L}\;,\quad
    n_b = \sigma_{e^+e^-\to e^+ e^-}^{\rm SM} \mathcal{L}\;,
\end{align}
and the integrated luminosity $\mathcal{L} = 5\abi$.
The resulting exclusion limits for different $m_{\Delta^{++}}$ are shown in dark blue in Fig.~(\ref{fig:bound},\ref{fig:bound3}) .

In addition to the total cross section, the analysis can be performed using differential distributions, as was done at LEP. To estimate the sensitivity, we rescale the existing limit on $f_R^{ee}/m_{\Delta^{++}}$ by a factor of $\sqrt{(\sigma_{\mathrm{LEP}}\mathcal{L}_{\mathrm{LEP}})/(\sigma_{e^+e^-}\mathcal{L}_{e^+e^-})}$, following Refs.~\cite{Dev:2018sel,Li:2024djp}. Here, $\sigma_{\mathrm{LEP}}$ and $\sigma_{\mathrm{e^+e^-}}$ denote the magnitudes of the NP contributions to the Bhabha scattering at LEP and CEPC/FCC-ee, respectively. The integrated luminosities are $\mathcal{L}_{\rm LEP} = 745~\text{pb}^{-1}$ and $\mathcal{L}_{e^+e^-} = 5\abi$.

\subsection{Muon colliders}

Given the current bound on the mass of $\dR$, we consider
the future muon collider (MuC) with 
the center-of-mass energy $\sqrt{s} = 3\tev\,(10\tev)$~\cite{MuonCollider:2022nsa,InternationalMuonCollider:2024jyv},
which provides opportunities for both precision tests of the process $\mu^+ \mu^- \to \mu^+ \mu^-$, and direct production of on-shell doubly-charged scalars via $\mu^+ \mu^- \to \Delta_R^{++}\Delta_R^{--}$, which are shown in Fig.~\ref{fig:mupmummumu} and Fig.~\ref{fig:deltadelta}, respectively.  

For the process $\mu^+ \mu^- \to \mu^+ \mu^-$, the expression of cross section is the same as that for the Bhabha scattering, with the substitution of $e$ to $\mu$. The benchmark cross section is shown in Fig.~\ref{fig:cross}.  
We consider the integrated luminosity $\mathcal{L} = 1\abi\, (10\abi)$ for 
$3\tev\, (10\tev)$ MuC, and baseline cut on the pseudo-rapidity of leptons $|\eta| < 2.5$~\cite{InternationalMuonCollider:2024jyv}. The constraint on $f_R^{\mu\mu}/m_{\Delta^{++}}$ is obtained from the precise measurement of the cross section for the process $\mu^+ \mu^- \to \mu^+ \mu^-$ with the exclusion limits derived using Eq.~\eqref{eq:crit}.

Besides, 
the 
pair production process $\mu^+\mu^-\to \Delta_R^{++}\Delta_R^{--}$ is possible if the center-of-mass energy $\sqrt{s}>2m_{\Delta^{++}}$.
The cross section is given by
\begin{align}
 \sigma_{\mu^+\mu^-\to \Delta_R^{++} \Delta_R^{--}} &= \frac{(f_R^{\mu\mu})^4 (s-2m_{\Delta^{++}}^2)}{64 \pi  s^2} \ln\frac{w_+}{w_-} \nn\\
 &\quad -\frac{(f_R^{\mu\mu})^4\sqrt{s(s-4m_{\Delta^{++}}^2)} }{32 \pi  s^2}\;,
\end{align}
with $w_{\pm } \equiv s-2m_{\Delta^{++}}^2 \pm \sqrt{s(s-4m_{\Delta^{++}}^2)}$.
The on-shell $\Delta_R^{\pm\pm}$ can subsequently decay into a pair of same-sign leptons with the same flavor. The partial decay width of $\Delta_R^{++}\to \ell^+\ell^+$, where $\ell = e,\mu$, is given by
\begin{eqnarray}
    \Gamma_{\ell\ell}=\frac{(f_R^{\ell \ell})^2 m_{\Delta^{++}}}{4\pi}\sqrt{1-\frac{m_\ell^2}{m_{\Delta^{++}}^2}}\;.
\end{eqnarray}

Since the sensitivity to the Yukawa couplings depends on the flavor combination of the two decay chains, we consider three distinct cases: $4e$, $4\mu$, and $2e2\mu$. From the collider simulation in Ref.~\cite{Belfkir:2023lot}, these final states are nearly background free, and the signal efficiencies are about $50\%-80\%$. 
We assume a conservative and universal signal efficiency of $ 50\%$, and derive the 95\% CL exclusion limits by requiring at least 3 signal events assuming no SM background~\cite{Junk:1999kv,Bhattiprolu:2020mwi}. The results for $3\tev$ and $10\tev$ MCs are presented in Fig.~\ref{fig:bound} and Fig.~\ref{fig:bound3}, respectively.

\subsection{\tf{$\mu$}{mu}TRISTAN}

By utilizing a low-emittance muon beam originally developed for the measurements of muon $g-2$
at J-PARC~\cite{Abe:2019thb}, a new collider design known as $\mu$TRISTAN has been proposed~\cite{Hamada:2022mua}.
By accelerating $\mu^+$ and $e^-$ beams up to $1\tev$ and $30\gev$, respectively,  the center-of-mass
energies of $2\tev$ and $346\gev$ can be achieved for colliding $\mu^+\mu^+$ and  $\mu^+e^-$, respectively.
At both colliders, lepton number violation (LNV) can be probed through same-sign dilepton signals arising from the Yukawa interactions given in Eq.~\eqref{eq:yuk}.

At the $\mu^+\mu^+$ collider, the \dcs $\dR$ can lead to the process $\mu^+ \mu^+ \to \ell^+ \ell^+$ for $\ell = e, \mu$, as depicted in Fig.~\ref{fig:mumumumu} and Fig.~\ref{fig:mumuee}. For $m_{\Delta_R^{++}} < 2 \tev$, the signal process $\mu^+ \mu^+ \to \gamma \Delta_R^{++}(\to \ell^+\ell^+)$ can also occur. The associated production of $\gamma$ and $\Delta_R^{++}$ is displayed in Fig.~\ref{fig:deltagamma1} and Fig.~\ref{fig:deltagamma2}, the latter of which involves the Yukawa interaction of $\Delta_R^{++}$. 
These processes were also studied within the type-II seesaw scenario in Ref.~\cite{Dev:2023nha,Das:2024kyk}.

On the other hand, the $\mu^+ e^-$ collider, which has been extensively utilized to search for CLFV signals, can also probe LNV through $t$-channel exchange of $\dR$ in the process $\mu^+ e^- \to \mu^- e^+$, see Fig.~\ref{fig:muemue}. In the following, we will investigate the sensitivities to the Yukawa couplings $f_R^{ee}$ and $f_R^{\mu\mu}$ in these processes.

Different from $\mu^+ \mu^- \to \mu^+ \mu^-$, the process $\mu^+ \mu^+ \to \ell^+ \ell^+$ receives contributions from $\dR$ via the $s$-channel, while SM $\gamma/Z$ exchange occurs in the $t$-channel. Using the notation of Eq.~\eqref{eq:xsec} with $i = \mu^+\mu^+\to \mu^+\mu^+$, the NP contribution from \dcs is given by 
\begin{align}
    \sigma_{\mu^+\mu^+ \to \mu^+\mu^+}^{\rm NP} = \left[\sigma_{\Delta}(t)+ \sigma_{\rm int}(t)\right]|^{t_{\rm max}}_{t_{\rm min}}\;,
\end{align}
while the expression of SM cross section is lengthy and omitted. In the above,  
\begin{align}
    \sigma_{\Delta} (t)&=\frac{(f_R^{\mu\mu})^4}{64\pi}\frac{t}{|D_\Delta (s)|^2} \;,\\
    \sigma_{\rm int}(t)&=\frac{(f_R^{\mu\mu})^2\alpha_{\rm em}}{4}\frac{s-m^2_{\Delta^{++}} }{(s-m_{\Delta^{++}}^2)^2+m_{\Delta^{++}}^2\Gamma_\Delta^2}\ln\frac{s+t}{t}
\nonumber\\
&\quad - \kappa_1 \tan^{-1}\frac{m_Z\Gamma_Z(s+2t)}{(m_Z^2+s+t)(m_Z^2-t)+m_Z^2\Gamma_Z^2}\nonumber\\
&\quad + \kappa_2 \ln\frac{(s+t+m_Z^2)^2+m_Z^2\Gamma_Z^2}{(t-m_Z^2)^2+m_Z^2\Gamma_Z^2}\;,
\end{align}
where $D_\Delta(s) \equiv s-m_{\Delta^{++}}^2 + i \Gamma_\Delta m_{\Delta^{++}}$, and we have introduced the definitions
\begin{align}
    \kappa_1 &=\frac{(f_R^{\mu\mu})^2\alpha_{\rm em}}{64c_W^2s_W^2} \frac{m_{\Delta^{++}} \Gamma_\Delta}{|D_\Delta (s)|^2}\;,\\
    \kappa_2 &=\frac{(f_R^{\mu\mu})^2\alpha_{\rm em}}{128c_W^2s_W^2}\frac{s-m_{\Delta^{++}}^2}{|D_\Delta (s)|^2}\;.
\end{align}
In the above, the minimal and maximal values of $t$ are defined as Eq.~\eqref{eq:tmin}. The infrared divergence appearing in $\sigma_{\rm int}$ for the Mandelstam variable $t \to 0$ or $t \to -s$ is regularized after imposing the pseudo-rapidity cut as that for the Bhabha scattering. The relation between $t$ and the pseudo-rapidity $\eta$ is given in Eq.~\eqref{eq:rapidity}. We require $|\eta|<2.5$, and obtain the cross section as depicted in Fig.~\ref{fig:cross}.

For the process $\mu^+ \mu^+ \to e^+ e^+$, the cross section from the $s$-channel exchange of $\Delta_R^{++}$  is given by
\begin{eqnarray}
&& \sigma_{\mu^+\mu^+ \to e^+e^+}= \frac{(f_R^{ee})^2 (f_R^{\mu\mu})^2}{128\pi}\frac{s}{|D_\Delta(s)|^2}\;.
\end{eqnarray}
Note that there is no SM contribution to this process, so that its cross section is several orders of magnitude smaller than that of $\mu^+\mu^+ \to \mu^+ \mu^+$, as illustrated in Fig.~\ref{fig:cross}.

The \dcs can also be produced in association with a photon $\mu^+ \mu^+ \to \gamma \Delta_R^{++}$, if the center-of-mass energy $\sqrt{s} > m_{\Delta^{++}}$. The decay $\Delta_R^{++}\to \ell^+ \ell^+$ for $\ell = e,\mu$ occurs subsequently.
The cross section for the associated production is
\begin{align}
\sigma_{\mu^+\mu^+ \to \gamma\Delta_R^{++}}&=\frac{e^2 (f_R^{\mu\mu})^2}{128 \pi s^2} \sigma_{\gamma\Delta}(t) |^{t_{\rm max}}_{t_{\rm min}}\;,
\end{align}
where 
\begin{align}
\sigma_{\gamma \Delta}(t)&=\frac{(s^2+m_{\Delta^{++}}^4)\ln(s-m_{\Delta^{++}}^2+t)}{s-m_{\Delta^{++}}^2}-4t\nonumber\\
&\qquad-\frac{(5s^2-4m_{\Delta^{++}}^2 s+m_{\Delta^{++}}^4)\ln t}{s-m_{\Delta^{++}}^2}\;.
\end{align}
The Mandelstam variable $t$ for this process is defined as
\begin{align}
    t=-\frac{s - m_{\Delta^{++}}^2}{1+e^{2\eta}}\;,
\end{align}
where $\eta$ denotes the pseudo-rapidity of the photon, 
and the minimal and maximal values of $t$ are given by
\begin{align}
    t_{\rm min}=-\frac{s - m_{\Delta^{++}}^2}{1+e^{-2\eta_m}}\, \quad
t_{\rm max}=-\frac{s - m_{\Delta^{++}}^2}{1+e^{2\eta_m}}\;.
\end{align}
In the analysis, we impose the cut on photon $|\eta| < \eta_m = 2.5$. 
From Fig.~\ref{fig:cross}, we can see that the cross section for $\mu^+\mu^+ \to \gamma\Delta_R^{++} $ increases with $m_{\Delta_R^{++}}$.

At the $\mu^+ e^-$ collider, the process $\mu^+ e^- \to \mu^- e^+$ can occur via $t$-channel exchange of $\Delta_R^{\pm\pm}$, which violates the lepton flavor number by two units $\Delta L_\mu = - \Delta L_e = 2$. 
The cross section is given by
\begin{eqnarray}
&\sigma_{\mu^+e^- \to \mu^-e^+ } = \dfrac{(f_R^{ee})^2(f_R^{\mu\mu})^2 }{128 \pi  s^2} \Big[s\dfrac{(s+2m_{\Delta^{++}}^2)}{s+m_{\Delta^{++}}^2}\nn\\
&\qquad\quad  -2m_{\Delta^{++}}^2 \ln\dfrac{s+m_{\Delta^{++}}^2}{m_{\Delta^{++}}^2}\Big]\;.
\end{eqnarray}
Since the flavor off-diagonal Yukawa couplings $f_R^{\alpha\beta}$ are smaller than the diagonal ones in order to satisfy the severe CLFV constraints, the contributions from the flavor off-diagonal couplings are not included.
The cross section $\sigma_{\mu^+e^- \to \mu^-e^+ }$ decreases with increasing $m_{\Delta^{++}}$, which is illustrated as the orange curve in Fig.~\ref{fig:cross}.

Similar to the analyses at future MuC, we evaluate the sensitivities of the processes depending on whether they have SM contributions.
For the process $\mu^+ \mu^+ \to \mu^+ \mu^+$,
which can occur in the SM,
we employ the statistical criterion of Eq.~\eqref{eq:crit}. For the processes $\mu^+ e^- \to \mu^- e^+$, $\mu^+ \mu^+ \to e^+ e^+$ and $\mu^+ \mu^+ \to \gamma \Delta_R^{++} (\to \ell^+ \ell^+)$,
which are purely beyond the SM, we require a minimum of three signal events, assuming negligible SM background. The resulting expected 95\% CL exclusion limits are presented in Fig.~\ref{fig:bound} and Fig.~\ref{fig:bound3}.

\section{Results and discussions}
\label{sec:discussion}

\begin{figure}[!t]
\centering
\includegraphics[width=.4\textwidth]{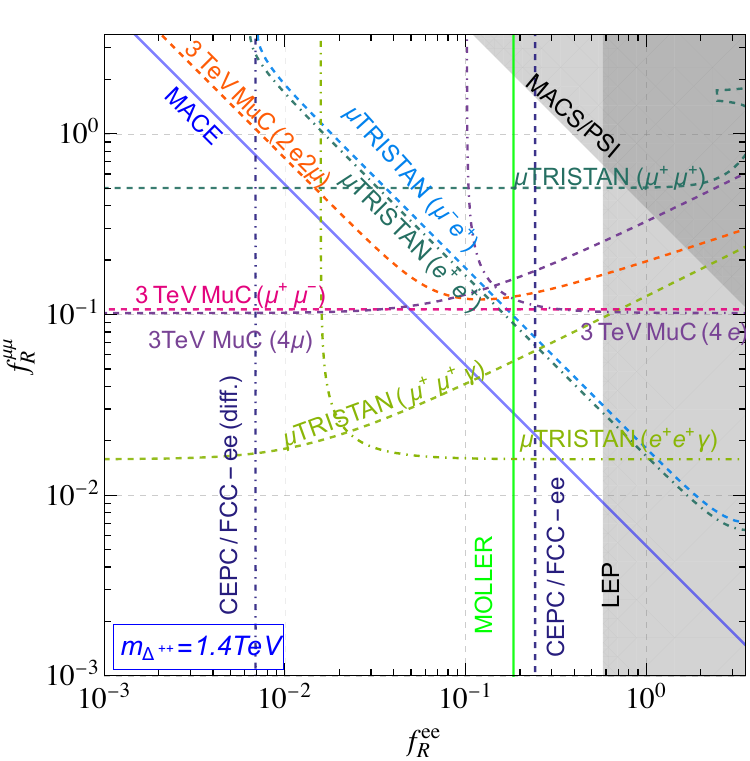}
\caption{ 
Current exclusion limits and future prospects of the Yukawa couplings $f_R^{ee}$ and $f_R^{\mu\mu}$ for $m_{\Delta^{++}} = 1.4~\text{TeV}$. Gray regions indicate current 95\% CL exclusion from measurements of $(g-2)_\mu$ ~\cite{Muong-2:2025xyk,Aliberti:2025beg}, Bhabha scattering at LEP~\cite{ALEPH:2013dgf}, and searches for $M-\overline{M}$ transition at MACS/PSI~\cite{Willmann:1998gd}. Projected 95\% CL sensitivity include those from future experiments: Bhabha scattering at CEPC/FCC-ee~\cite{CEPCStudyGroup:2018ghi,FCC:2018evy}, Møller scattering at MOLLER~\cite{MOLLER:2014iki}, $M-\overline{M}$ transition at MACE~\cite{Bai:2024skk}. Also shown are prospects from high-energy analogues at a 3 TeV MuC~\cite{MuonCollider:2022nsa,InternationalMuonCollider:2024jyv} and $\mu$TRISTAN with $\mu^+\mu^+/\mu^+e^-$ beams~\cite{Hamada:2022mua}, along with limits from direct pair and associated production processes at the MuC.
The constraint from muon $g-2$ is too weak and not shown.
}
     \label{fig:bound}
\end{figure}

\begin{figure}[!t]
\centering
\includegraphics[width=.4\textwidth]{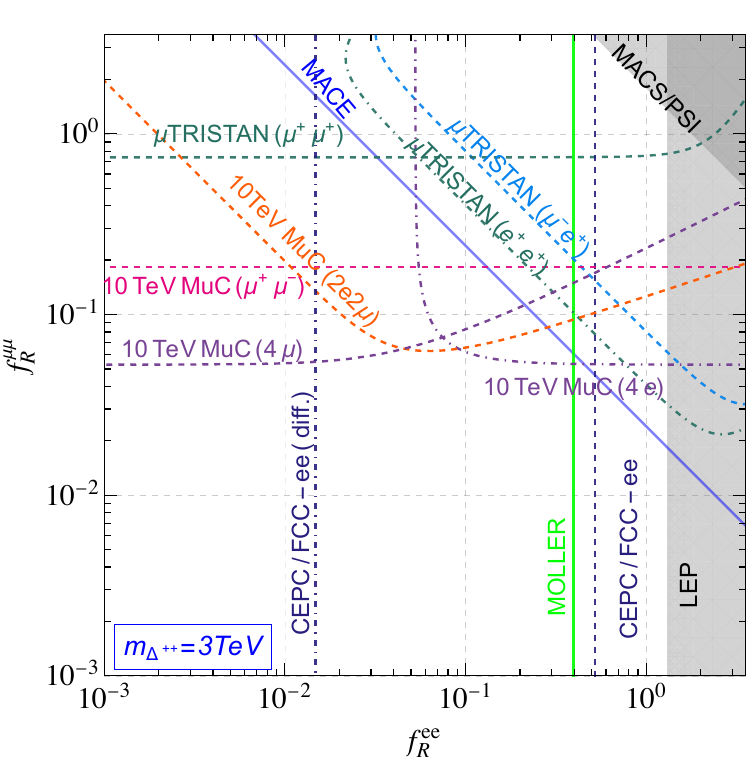}
	\caption{ Same as Fig.~\ref{fig:bound}, but for $m_{\Delta^{++}} = 3$~TeV and 10~TeV MuC. 
    }
     \label{fig:bound3}
\end{figure}

In this section, we discuss the combined constraints on the Yukawa couplings $f_R^{ee}$ and $f_R^{\mu\mu}$ from the low-energy observables and high-energy processes. Two benchmark masses of the doubly-charged scalar, $m_{\Delta_R^{++}} = 1.4\ \text{TeV}$ and 3 TeV satisfying the LHC bound~\cite{ATLAS:2022pbd}, are considered in Fig.~\ref{fig:bound} and Fig.~\ref{fig:bound3}, respectively.  
Processes at MuC and $\mu$TRISTAN are labeled by their final states, which are given in parentheses. For the processes involving on-shell $\dR$ at colliders, we assume $f_R^{\tau\tau} =0$ to derive its decay branching ratios.

Bhabha and Møller scattering are sensitive to $f_R^{ee}$. For $m_{\Delta_R^{++}} = 1.4\ \text{TeV}$ (3 TeV), LEP data~\cite{ALEPH:2013dgf} exclude the region $f_R^{ee} > 0.58\ (1.23)$, while the future MOLLER experiment~\cite{MOLLER:2014iki} will be able to rule out $f_R^{ee} > 0.18\ (0.39)$. At CEPC/FCC-ee~\cite{CEPCStudyGroup:2018ghi,FCC:2018evy}, Bhabha scattering cross-section measurements can reach a sensitivity of $f_R^{ee} > 0.24\ (0.51)$. This can be significantly improved with differential cross-section measurements, testing values of $f_R^{ee}$ as low as $0.007 \ (0.015)$.

Similarly, the processes $\mu^+ \mu^- \to \mu^+ \mu^-$ and $\mu^+ \mu^+ \to \mu^+ \mu^+$  are able to probe $f_R^{\mu\mu}$.  
For $m_{\Delta_R^{++}} = 1.4\ \text{TeV}$ (3 TeV), a 3~TeV MuC~\cite{MuonCollider:2022nsa,InternationalMuonCollider:2024jyv} can test $f_R^{\mu\mu} > 0.1$ via precise measurements of $\mu^+ \mu^- \to \mu^+ \mu^-$, surpassing current limits from $(g-2)_\mu$~\cite{Muong-2:2025xyk,Aliberti:2025beg}. This sensitivity, however, diminishes at a higher center-of-mass energy, $\sqrt{s} = 10\ \text{TeV}$. Although $\mu^+ \mu^+ \to \mu^+ \mu^+$ at a 2~TeV $\mu$TRISTAN has a larger cross section than $\mu^+ \mu^- \to \mu^+ \mu^-$ at a 3~TeV MuC (see Fig.~\ref{fig:cross}), its expected constraint is weaker due to the assumed lower integrated luminosity of $12\ \text{fb}^{-1}$.

The $M-\overline{M}$ transition excludes the upper-right region in the $(f_R^{ee}, f_R^{\mu\mu})$ plane. In particular, MACS/PSI~\cite{Willmann:1998gd} has ruled out $f_R^{ee} f_R^{\mu\mu} > 0.39$ (1.78) for $m_{\Delta_R^{++}} = 1.4\ \text{TeV}$ (3 TeV)\,\footnote{It is noted that we require $f_R < \sqrt{4\pi}$ in the exclusion plots for benchmark illustration. For perturbative consistency up to the $D$-parity breaking scale, a more conservative requirement is $f_R < 1$~\cite{Sahu:2006pf}.} The lower limit can be extended by the future MACE experiment~\cite{Bai:2024skk} to $f_R^{ee} f_R^{\mu\mu} > 3\times 10^{-3}$ $(2.4\times 10^{-2})$.  
As a high-energy counterpart, $\mu^+ e^- \to \mu^- e^+$ at $\mu$TRISTAN is able to probe the region of $f_R^{ee} f_R^{\mu\mu} < 0.02$ (0.08), improving upon the MACS/PSI bounds by more than one order of magnitude.

Direct production of the \dcs offers another avenue to simultaneously constrain $f_R^{ee}$ and $f_R^{\mu\mu}$. At a MuC, the pair production $\mu^+ \mu^- \to \Delta_R^{++} \Delta_R^{--}$ depends on $f_R^{\mu\mu}$, while the decay branching ratios of $\Delta_R^{\pm\pm} \to e^\pm e^\pm, \mu^\pm \mu^\pm$ vary with $f_R^{ee}$ and $f_R^{\mu\mu}$.  
We find that $\mu^+\mu^- \to \Delta_R^{++} \Delta_R^{--} \to 4e$ is highly sensitive to both couplings. For $m_{\Delta_R^{++}} = 1.4\ \text{TeV}$ (3~TeV) at a 3 TeV (10~TeV) MuC, this channel can exclude much of the parameter space where $f_R^{ee}, f_R^{\mu\mu} > 0.1\ (0.05)$.  
In comparison, $\mu^+\mu^- \to \Delta_R^{++} \Delta_R^{--} \to 4\mu$ tests $f_R^{\mu\mu}$ down to $0.1\ (0.05)$ when $f_R^{ee} < 0.05$; for larger $f_R^{ee}$, sensitivity to $f_R^{\mu\mu}$ drops quickly.  
The process $\mu^+\mu^- \to \Delta_R^{++} \Delta_R^{--} \to 2e2\mu$ exhibits a turnover in sensitivity: it increases with $f_R^{ee}$ up to about $0.1\ (0.04)$, beyond which it declines.

A similar finding holds for the associated production at $\mu$TRISTAN. The channel $\mu^+\mu^+ \to \gamma \Delta_R^{++} (\to e^+ e^+)$ is highly sensitive to both couplings, excluding most of the region where $f_R^{ee}, f_R^{\mu\mu} > 0.018$ for $m_{\Delta_R^{++}} = 1.4\ \text{TeV}$.  
Meanwhile, $\mu^+\mu^+ \to \gamma \Delta_R^{++} (\to \mu^+ \mu^+)$ can test $f_R^{\mu\mu}$ down to 0.018 if $f_R^{ee} < 6 \times 10^{-3}$, with sensitivity decreasing for larger $f_R^{ee}$. Since the center-of-mass energy of $\mu$TRISTAN with $\mu^+\mu^+$ beams is 2~TeV, the associated production process is not applicable for $m_{\Delta_R^{++}}=3\tev$. 

Besides, the process $\mu^+ \mu^+ \to e^+ e^+$, mediated by an off-shell $\Delta_R^{++}$ at $\mu$TRISTAN, probes the product of the two Yukawa couplings. Its sensitivity relative to $\mu^+ e^- \to \mu^- e^+$ at $\mu$TRISTAN depends crucially on the mass of $\dR$. 
For $m_{\Delta_R^{++}}=1.4\tev$, $\mu^+e^- \to \mu^-e^+$ provides a comparable expected exclusion, while $\mu^+ \mu^+ \to e^+ e^+$ becomes superior for $m_{\Delta_R^{++}}=3\tev$.

Thus far, we have assumed $f_R^{\tau\tau} = 0$. Introducing a non-zero value leaves the sensitivities of low-energy probes and high-energy precision measurements in Fig.~\ref{fig:feynman} (a), (b), (d), (e), and (h) unchanged. In contrast, the reach of direct production processes at MuC and $\mu$TRISTAN in Fig.~\ref{fig:feynman} (c), (f), (g) is diminished, as a non-zero $f_R^{\tau\tau}$ reduces the decay branching ratios of $\dR$ into electrons and muons.
Moreover, by measuring the invariant-mass distributions constructed from the tau decay products in the process $pp \to \Delta_R^{++} \Delta_R^{--} \to \ell^\pm \ell^\pm \tau^\mp \tau^\mp$, which encode the polarization information of the tau leptons~\cite{Sugiyama:2012yw}, it is possible to probe the chiral structure of the Yukawa interaction in Eq.~\eqref{eq:yuk}. Such observables can therefore be used to experimentally distinguish $\Delta_R^{\pm\pm}$ from $\Delta_L^{\pm\pm}$.

\section{Conclusion}
\label{sec:conclusion}

In this work, we have investigated the phenomenology of a TeV-scale \dcs in various low-energy and high-energy experiments. 
Such a \dcs arises naturally from the left-right symmetric model with $D$-parity breaking. Even with the stringent constraints from charged lepton flavor violation searches, the flavor-diagonal couplings of the right-handed doubly-charged scalar $\Delta_R^{\pm\pm}$ to leptons can be sizable.

We focus on the region of Yukawa couplings $f_R^{ee}, f_R^{\mu\mu} \gtrsim 10^{-3}$ for the mass of \dcs being around $\mathcal{O}(1)\tev$. Once being observed, it would provide compelling indirect evidence for the type-I seesaw mechanism as the origin of neutrino masses, particularly in the absence of detected CLFV signals.

We study the contributions of $\Delta_R^{\pm\pm}$ to the low-energy observables, including the parity-violation asymmetry in Møller scattering, muon $ g-2$, and muonium-antimuonium transition probability, and provide the full analytical expressions for the cross sections of the processes involving the \dcs $\dR$ at future lepton colliders (cf. Fig.~\ref{fig:feynman}).

We derive constraints on the couplings $f_R^{ee}$ and $f_R^{\mu\mu}$ for $m_{\Delta_R^{++}} = 1.4\ \text{TeV}$ and $3\ \text{TeV}$ by analyzing both low- and high-energy probes at their projected sensitivities, as shown in Fig.~\ref{fig:bound} and Fig.~\ref{fig:bound3}, respectively. Our results show that future measurements of the parity-violating asymmetry in Møller scattering (MOLLER experiment) and searches for $M-\overline{M}$ transition (MACE experiment) can significantly extend the current bounds on the Yukawa couplings $f_R^{ee}$ and $f_R^{\mu\mu}$. Furthermore, high-energy lepton colliders could probe coupling regions of $f_R^{ee} \gtrsim 10^{-1}$ and $f_R^{\mu\mu} \gtrsim 10^{-2}$ using the total cross sections. Sensitivity to $f_R^{ee}$ could be improved to the level of $\sim 10^{-2}$ by analyzing the differential distribution of Bhabha scattering at CEPC/FCC-ee.

\appendix

\bigskip

\begin{acknowledgments}

We would like to thank Jian Tang and Yongchao Zhang for helpful discussions.
GL is supported by the National Natural Science Foundation of China under Grants No.~12347105 and No.~12505127, and the Guangdong Basic and Applied Basic Research Foundation (2024A1515012668). Jin Sun is supported by IBS under the project code, IBS-R018-D1.		

\end{acknowledgments}


\bibliographystyle{apsrev4-1}

\bibliography{reference}


\end{document}